\documentclass[aps,twocolumn]{revtex4}
\usepackage{psfrag}

\usepackage{graphicx}
\usepackage{amsmath}
\usepackage{amsfonts}
\usepackage{amssymb}
\usepackage{color}
\usepackage{subfigure}

\begin{document}

\title{\textbf{Vortex shedding patterns, their competition, and chaos in flow past inline oscillating rectangular cylinders}}
\date{\today}
\author{Srikanth T$^1$, Harish N Dixit$^{1}$ \footnote{Presently at the University of British Columbia, Vancouver, Canada}, Rao Tatavarti$^2$, and Rama Govindarajan$^1$ }
\email{rama@jncasr.ac.in}
\affiliation{1. Engineering Mechanics Unit, Jawaharlal Nehru Centre for Advanced
Scientific Research, Jakkur, Bangalore 560064, India \\
2. Department of Civil Engineering, Gayatri Vidya
Parishad College of Engineering, Madhurawada, Visakhapatnam 530048, India.
}

\begin{abstract}
The flow past inline oscillating rectangular cylinders is studied numerically at a Reynolds number representative of two-dimensional flow. A symmetric mode, known as S-II, consisting of a pair of oppositely-signed vortices on each side, observed recently in experiments, is obtained computationally. A new symmetric mode, named here as S-III, is also found. At low oscillation amplitudes, the vortex shedding pattern transitions from antisymmetric to symmetric smoothly via a regime of intermediate phase. At higher amplitudes, this intermediate regime is chaotic.  The finding of chaos extends and complements the recent work of Perdikaris {\em et al.} \cite{perdikaris}. Moreover it shows that the chaos results from a competition between antisymmetric and symmetric shedding modes. Rectangular cylinders rather than square are seen to facilitate these observations. A global, and very reliable, measure is used to establish the existence of chaos.
\end{abstract}

\maketitle

\section{Introduction}
Vortex shedding from bluff bodies is an extensively studied problem. The preferred mode of vortex shedding, in the uniform flow past a fixed body, is antisymmetric. On the other hand a body forced to oscillate in quiescent fluid would be expected to shed a symmetric pattern of vortices. Thus, for a body oscillating inline in a uniform external flow, as the frequency of oscillation is increased with all other parameters remaining fixed, we may expect a transition from antisymmetric to symmetric shedding. Both kinds of shedding have been observed \cite{rockwell,balabani,zhou}. Secondly, it has been seen experimentally that there is more than one kind of symmetric shedding \cite{xu}. Our objective here is to improve our understanding of the frequency response of the system, in terms of the spatial arrangements of vortices and the transitions therein. Studies on such flows have found application in predicting the loading on offshore structures \cite{williamson}. Also, this is a simple example of the flow due to an accelerating body. The prediction of flow patterns in its wake can be important in various contexts, such as in the tracking of underwater bodies.

Griffin \& Ramberg \cite{ramberg} were among the first to study vortex shedding from an inline oscillating circular cylinder in a freestream. They found that the vortex shedding frequency locks on to the frequency of cylinder oscillation for $1.2 \leq f_e/f_o \leq 2.5$, where $f_e$ is the frequency of cylinder oscillation and $f_o$ would have been the frequency of vortex shedding if the cylinder were held stationary. The subscripts $e$ and $o$ have been chosen to stand for `excitation' and `original' respectively. Both primary lock-on, where the shedding frequency $f_s=f_e$, and subharmonic lock-on, with $f_s=f_e/2$ were observed. Ongoren \& Rockwell \cite{rockwell} carried out experiments with a circular cylinder oscillating at an angle $\alpha$ with a uniform freestream. Outside the lock-on regime, competition between symmetric and antisymmetric modes in the form of switching of modes in a single experiment were observed. In some recent experiments Konstantinidis \& Balabani \cite{balabani} too noted the symmetric mode mentioned above, where all vortices shed from the top wall were of one sign, while those shed from the bottom wall were of the opposite sign. This pattern is called the S-I mode of shedding. In another experimental study on a circular cylinder, Xu \textit{et al.} \cite{xu} discovered a new mode of symmetric shedding, which they named S-II. Two vortices of opposite sense were shed from each side (top and bottom) during each cycle. This mode was observed for high frequencies and amplitudes. There was considerable reverse flow during a part of the cycle, which aided in the formation of opposite signed vortices on a given side of the cylinder. Very few numerical studies have reported the symmetric S-I shedding, Zhou \& Graham's \cite{zhou} being one. To our knowledge, the S-II mode has not been found numerically before.

Besides systematic shedding, we could have chaotic shedding. Chaos in flow around an inline oscillating circular cylinder (or equivalently, in oscillating flow past a fixed cylinder) was reported by Vittori \& Blondeaux \cite{vittori} and Perdikaris \textit{et al.} \cite{perdikaris}. The former study had no mean flow, and showed that the route to chaos is quasiperiodic, and the latter study attributed chaos to mode competition. However no evidence of mode competition was provided. One goal of the present study is to present direct evidence of competition between antisymmetric and symmetric shedding, and the resulting chaos.  Ciliberto \& Gollub \cite{gollub1,gollub2} showed that competition between different modes in parametrically forced surface waves can result in chaos. Their study also revealed the existence of chaotic `windows'. They remarked that this could be a common cause of chaos in systems in which different spatial structures can exist. The present flow is shown to be one example.

In a circular or square cylinder, the behaviour of each row of shed vortices can be clouded by interaction with the opposite row. For this reason, we study the flow past rectangular cylinders of various aspect ratios. To our knowledge, there have been no previous studies on vortex shedding from inline oscillating rectangular cylinders. We show the existence of symmetric S-I and S-II modes, and a new S-III mode, besides the Couder-Basdevant mode. We then discuss a physical mechanism for the enhancement of S-II shedding from rectangular cylinders. We show that the transition from antisymmetric to symmetric shedding, as $f_e$ is increased, for lower oscillation amplitudes occurs via periodic flows of different phase. At higher oscillation amplitudes we find windows of chaos between the regimes where shedding modes are antisymmetric and symmetric.

\section{Problem formulation}

The numerical procedure described in Dixit \& Babu \cite{harish} was employed after making suitable changes. Flow solutions are obtained by solving the Lattice-Boltzmann equation

\begin{equation}
\frac{\partial f_i}{\partial t} + e_{ik} \frac{\partial f_i}{\partial x_k} = \frac{1}{\tau} (f^{eq}_i - f_i).
\end{equation}

As in the usual nomenclature, $f_i(\textbf{X},t)$ and $f^{eq}_i(\textbf{X},t)$ are the nonequilibrium and equilibrium distribution functions in $i^{th}$ direction, $\tau$ is the time between two successive collisions. Thus the density $\rho = \Sigma_i f_i$ and the momentum vector is $\rho u_k = \Sigma_i f_i e_{ik}$. No-slip on walls is imposed by employing the bounce-back scheme. 

It is straightforward to show that in the absence of rotation of the cylinder, the vorticity and continuity equations in a cylinder-fixed frame are the same as those in the lab-fixed frame. In the cylinder-fixed coordinate system we have an oscillating inlet flow \cite{rodi}, $u_x = U_{\infty} + A \Omega \sin{\Omega t}$, $A$ is the amplitude of the displacement of the body, $\Omega$ ($ \equiv 2 \pi f_e$) is the frequency of oscillation, $u_x$ and $u_y$ are the streamwise and transverse components of fluid velocity respectively. The other boundary conditions are ${\partial u_x}/{\partial x} = {\partial u_y}/{\partial x} = 0$ at the outlet; $u_x = u_y = 0$ on the cylinder; $u_x = U_{\infty} + A \Omega \sin{\Omega t}$, $u_y = 0$ at the top and bottom surfaces.
The characteristic length in this study is taken to be the height $D$ of the body, and the characteristic velocity is $U_{\infty}$, so the Reynolds number is defined as $Re \equiv U_{\infty} D/\nu$, and the Strouhal number as $St \equiv f_s D/U_\infty$, $f_s$ being the shedding frequency of the vortices.

\begin{figure}[h]
%\centering
\subfigure{
\label{fig:domain}
\includegraphics[scale=0.2]{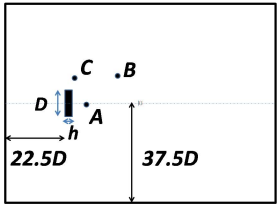}}
\subfigure{
\label{fig:grid}
\includegraphics[scale=0.15]{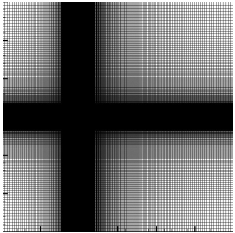}} 
\caption{(a) Schematic of the domain used in the simulations. Time signals are stored at the  monitor points \textit{A, B} and \textit{C}, located at (2$h$,0), ($h$, $D$) and (2.25$D$, 0.5$D$) respectively (not to scale). (b) The grid used.}
\end{figure}

Figure \ref{fig:domain} shows a part of the domain used for the simulations. The size of the domain is $75D \times 75D$ with nonuniform grids of up to a million grid points, whose arrangement is shown in figure \ref{fig:grid}. The values for the domain size and number of grid points were arrived at after doing a grid independence study. The reason for the use of a large domain is twofold: to minimize the effects of outlet boundary conditions on the wake and to ensure that the symmetric modes are insensitive to the lateral extent of the domain. All simulations in this study were carried out at $Re = 200$. 

The numerical approach was validated as follows. Strouhal numbers obtained from flow past a fixed square cylinder were found to be in good agreement for a range of Reynolds numbers from $50$ to $250$ with those from Okajima's experiments \cite{okajima} and Ansumali \textit{et al.}'s numerical simulations \cite{ansumali}. With an oscillating square cylinder, the dominant frequencies in the spectrum for the wall normal velocity $u_y$ at a typical location obtained from our simulations was found to be in good agreement with the spectrum for the lift coefficient obtained by Minewitsch \textit{et al.} \cite{rodi}. Further details are available in Srikanth \textit{et al.} \cite{srikanth}. Next, as done in \cite{rodi}, the frequency ratio was fixed at $f_e/f_o = 1.6$ and $A/D$ was varied from $0.15$ - $0.4$ in steps of $0.05$. In excellent agreement with \cite{rodi}, the lock-on window between $0.15 \leq A/D \leq 0.4$, where $f_s=0.5 f_e$ is reproduced.
 
\begin{figure}[h]
\centering
\subfigure[]{
\label{fig:couder}
\includegraphics[scale=0.18]{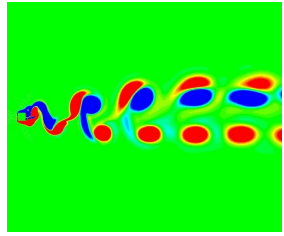}} 
\subfigure[]{
\label{fig:xu}
\includegraphics[scale=0.18]{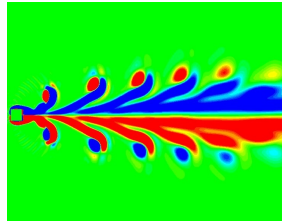}}
\caption{Qualitative comparison of shedding modes for the square cylinder. Vorticity fields are shown. (a) $f_e/f_o=2$ \& $A/D = 0.175$. This mode is similar to the one obtained by Couder \& Basdevant \cite{couder}. (b) $f_e/f_o= 1.73$ \& $A/D = 0.5$. This is S-II mode of shedding, similar to that seen in the experiment of Xu {\em et al.} \cite{xu}.}
\label{validation}
\end{figure}

For $Re = 200$, $A/D = 0.175$ and $f_e/f_o = 2$ in the square cylinder case, the vorticity field qualitatively matches with one of the experimental results of Couder \& Basdevant \cite{couder}. This mode, shown in figure \ref{fig:couder}, consists of two rows: one with binary vortices and the other with single vortices. We are able to obtain the S-II mode of vortex shedding behind a square cylinder, figure \ref{fig:xu}, for $Re = 200$, $A/D = 0.5$ and $f_e/f_o = 1.73$. Xu \textit{et al.} \cite{xu} were the first to report the S-II mode of shedding experimentally in the case of a circular cylinder. The amplitude ratio and frequency were the same in their experiment as the values used here, but their Reynolds number was 500. To the best of our knowledge, the S-II and Couder-Basdevant modes have not been seen computationally before.

\section{Results}

\subsection{Modes of vortex shedding}
\begin{figure}[]
\centering
%\subfigure[$f_e/f_o = 0.5$]{
%\label{fig:f05_v}
%\includegraphics[scale=0.135]{asp4_amp01_f05.jpg}} 
%
%\subfigure[$f_e/f_o = 1.5$]{
%\label{fig:f15_v}
%\includegraphics[scale=0.135]{asp4_amp01_f15.jpg}}
\subfigure[$f_e/f_o = 2$]{
\label{fig:f2_v}
\includegraphics[scale=0.18]{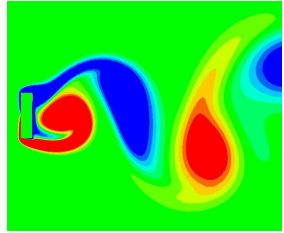}}
\subfigure[$f_e/f_o = 3$]{
\label{fig:f3_v}
\includegraphics[scale=0.18]{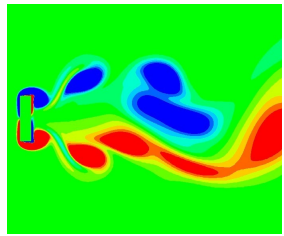}} 
\subfigure[$f_e/f_o = 4$]{
\label{fig:f4_v}
\includegraphics[scale=0.18]{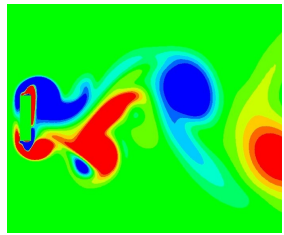}} 
\subfigure[$f_e/f_o = 5$]{
\label{fig:f5_v}
\includegraphics[scale=0.18]{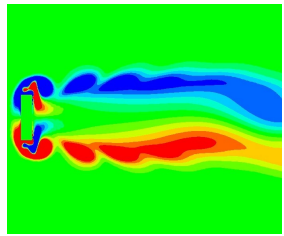}}
\caption{Vorticity fields at a typical time for $A/D=0.1$ at various excitation frequencies for a cylinder of aspect ratio $4$.}
\label{fig:amp01_v}
\end{figure}

When the oscillation frequency $f_e$ is very small, the flow is not too different from that past a fixed cylinder, except that the Reynolds number now is slowly varying. One expects, and finds, a slightly modified Karman street behind the body in this case. The same is true when the amplitude of oscillation $A/D$ is small, since the oncoming flow merely sees a slightly modified body on an average. To observe competition between symmetric and antisymmetric shedding, one needs an effective oscillation Reynolds number $Re_o= (\Omega A) D/\nu$ which is not negligible compared to that of the incoming flow. Although we carried out many simulations to ensure that our results are general, we present only a few typical ones. The shedding pattern changes as the excitation frequency $f_e$ is increased from $0.5 f_o$, half the natural shedding frequency of a stationary cylinder, to five times this value, some examples are shown in figure \ref{fig:amp01_v}. A rectangular cylinder of aspect ratio $4$ is used here and $A/D$ is fixed at $0.1$. At low $f_e$ the shedding is antisymmetric, and goes to symmetric shedding as $f_e$ increases. At $f_e/f_o=5$ we have the symmetric S-I mode, with all the top vortices being of one sign, and all the bottom of the opposite sign. At moderate $f_e/f_o$, the shedding is neither symmetric nor antisymmetric, but the upper and lower vortices are shed with a phase between $0$ and $\pi$ (or $\pi$ and $2\pi$). The flow however is still periodic. In some cases, vortex merger on each side of the cylinder is promoted, and the pattern downstream becomes antisymmetric. At small $f_e$, the shedding frequency $f_s$ is close to $f_o$. However, as $f_e/f_o$ is increased beyond $2$, $f_s$ decreases before locking on to a subharmonic of $f_e$, and then starts increasing proportionately with $f_e$, such that $f_s/f_e = 0.25$ for $3 \leq f_e/f_o \leq 4$. Beyond this range $f_s$ steadily decreases with further increase in $f_e$. The lock-on is similar to those seen on circular cylinders \cite{balabani}. Note that at these high frequencies shedding occurs on a given surface once every four complete oscillations, rather than once in every other oscillation.

\begin{figure}[]
\centering
\includegraphics[scale=0.35]{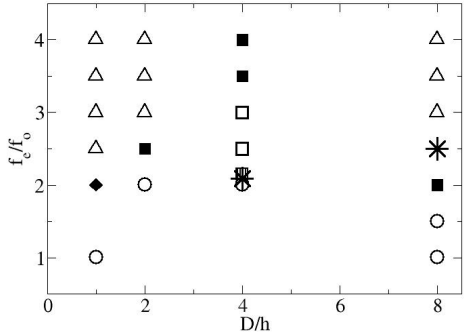} 
\caption{Flow patterns in the wake of an incline oscillating rectangular cylinder at $Re=200$ and $A/D=0.175$. Circles: antisymmetric shedding, squares: symmetric shedding. The solid squares indicate the S-II mode, the open squares stand for the S-I mode, while the patterned square indicates an S-III shedding. Triangles: mixed mode, where the shedding is symmetric but the vortices arrange themselves into an antisymmetric pattern downstream. Stars: chaotic flow,  single solid diamond: the Couder-Basdevant mode.}
\label{fig:aspect}
\end{figure} 

Next, choosing $A/D=0.175$, we summarise in figure \ref{fig:aspect} the patterns of vortex shedding observed on cylinders with aspect ratio $D/h = 1, 2, 4 $ and $8$. This higher oscillation amplitude will be seen to contrast with the lower $A/D$ discussed above, in particular in the transition from an antisymmetric pattern of shedding to a symmetric. In line with our expectations, it is easier to observe symmetric shedding behind a rectangular cylinder rather than a square one. The symmetric modes obtained may be classified into three types, S-I to S-III. As mentioned earlier, the first two have been observed in experiments before, but on circular cylinders \cite{rockwell,xu}. The letter S signifies a symmetric pattern, while the number denotes how many pairs of shed vortices may be associated with one time period of the flow. Thus the cylinder sheds one vortex of each sign both at the top and the bottom in an S-II mode. At higher oscillation frequencies, the flow displays what we term as a mixed mode. 

\begin{figure}[h]
\includegraphics[scale=0.18]{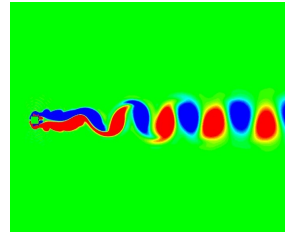} 
\caption{Mixed mode in the case of square cylinder, $f_e/f_o=4$, $A/D = 0.175$. } 
\label{fig:mixedmode}
\end{figure}

To discuss the mixed mode, shown in figure \ref{fig:mixedmode}, we choose a square geometry. Such a pattern has also been seen by Konstantinidis \& Balabani \cite{balabani}. The shedding off the cylinder is actually symmetric, but some distance downstream, the shed vortices arrange themselves in an antisymmetric pattern, much like a Karman street, but with a larger spacing, and a correspondingly lower Strouhal number of 0.92 times that of a fixed square cylinder at this Reynolds number. The oscillating square cylinder together with the symmetric portion of its wake corresponds roughly to a stationary body with an effective $D/h$ less than $1$. In fact the Strouhal number of the downstream portion of this figure is the same as that of a body whose aspect ratio is $0.67$. In taller geometries, the mixed mode is actually encouraged to occur by merger events of vortices of one sign, some of which are evident in figures \ref{fig:f3_v} and \ref{fig:f4_v}. The downstream behaviour again becomes antisymmetric. With all other parameters held constant, and reducing $h$ alone, i.e., using a taller rectangular cylinder rather than a square, we would reduce the relative size of the boundary layer and therefore the strength of the shed vortex. The pressure oscillations, which normally promote antisymmetric shedding, are correspondingly reduced, and so the symmetric pattern should persist further downstream for a given oscillation frequency. This is indeed manifested (not shown).

\begin{figure}[h]
\centering
\subfigure[]{
\label{fig:s3}
\includegraphics[scale=0.18]{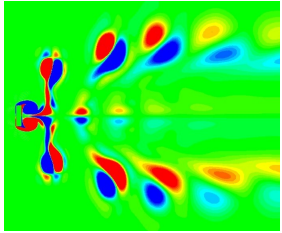}} 
\subfigure[]{
\label{fig:s2}
\includegraphics[scale=0.18]{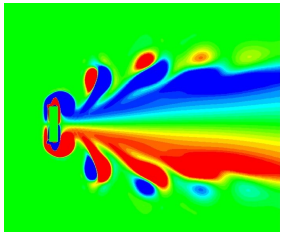}}
\caption{Two of the modes of shedding at $A/D = 0.175$ on a body of aspect ratio $4$. (a) The S-III mode at $f_e/f_o = 2.15$. Three pairs of binary vortices are shed. (b) The S-II mode at $f_e/f_o = 4$. In this mode two binary vortices are shed during each time period.}
\label{fig3}
\end{figure}

Returning to our discussion on the aspect ratio of $4$, the wake pattern changes from a Karman street, followed by a chaotic pattern, through S-III and then S-II, followed by S-I with increase in the frequency of oscillation. The S-III mode, shown in figure \ref{fig:s3}, is simply the S-II mode with an extra pair of vortices appearing close to the centreline. It is classified separately since it appears on the other side of chaos in the transition from antisymmetric shedding. A sample of S-II shedding on this cylinder is shown in figure \ref{fig:s2}.

\begin{figure}[h]
\centering
\subfigure[]{
\includegraphics[scale=0.25]{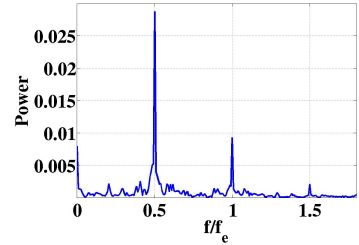}}
\subfigure[]{
\includegraphics[scale=0.25]{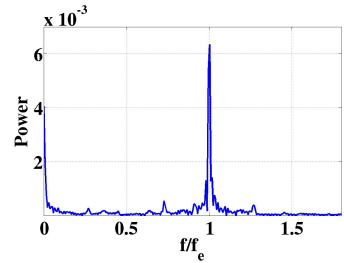}}
\caption{Power spectra at the monitor point $A$ for $D/h =4$ and $A/D = 0.175$. (a) Subharmonic shedding at $f_e/f_o = 2 $. (b) The shedding is harmonic (symmetric) for $f_e/f_o = 2.15$.}
\label{fig:subharm}
\end{figure}

The dominant frequency for $f_e/f_o = 2$ is $f_s = f_o$ (figure \ref{fig:subharm}). This is indicative of subharmonic lock-on \cite{barbi,ramberg}, while the shedding is locked on to the oscillation of the cylinder in the S-III mode.

%In Fig. \ref{fig:antisymm_phase} the frequency measured for vorticity, $\omega$, is half the excitation frequency. One vortex is shed during a time period of inflow oscillation. In terms of cylinder oscillation both anticlockwise and clockwise vortices are shed when the cylinder is moving to the right extreme and during this $Re_{effective} > 200$ ($Re_{effective}$ is defined as the $Re$ based on the instantaneous inlet velocity, i.e. $Re_{effective} = U(t) D/ \nu $).

\subsection{Mechanism for S-II mode}

\begin{figure}[]
\centering
\includegraphics[scale=0.3]{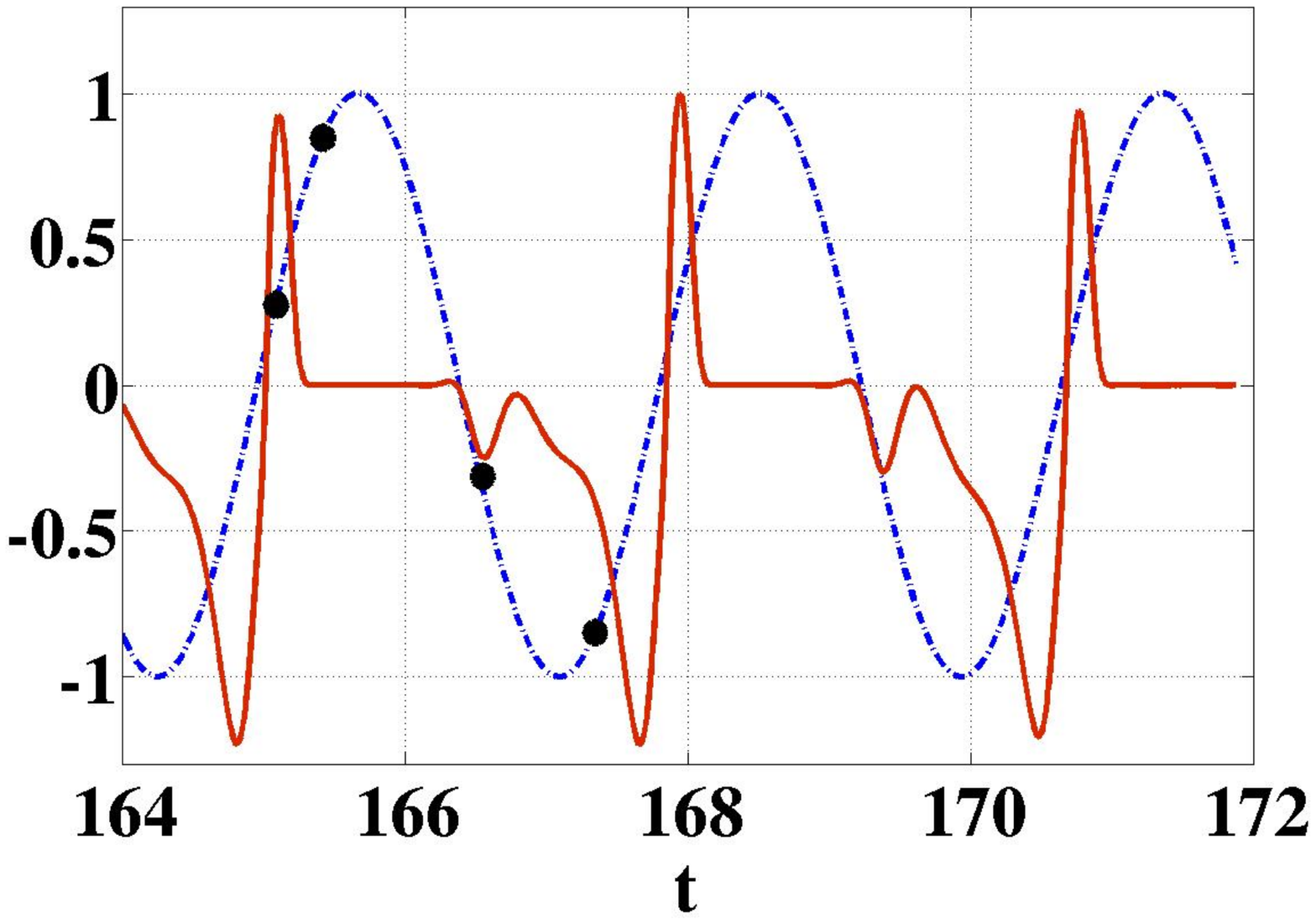}
\caption{Phase information for the S-II mode for $D/h = 8$. Solid red line: the vorticity $\omega$ at monitor point B. Blue dashed line: inlet velocity $U_{total}$. Here $f_e/f_o = 2$  and $A/D = 0.175$. The circles indicate the phases at which the vorticity field is shown in figures \ref{fig:p1} - \ref{fig:p4}.}
\label{fig:s2_phase}
\end{figure}

Figure \ref{fig:s2_phase} shows the time signal of the vorticity $\omega$ at the monitor point C. In the absence of a mean flow, it is easy to visualise the alternate shedding of oppositely signed vortices when the cylinder is moving to and fro. The mean flow advects both vortices downstream. This S-II mode of shedding is aided by the `ground effect'. The primary vortices accelerate the fluid in the wake region towards the cylinder, and due to the larger area available on a rectangle rather than a square at the lee surface, significant vorticity of opposite sign is generated. This effect is similar to the one studied by Carnevale \textit{et al.} \cite{orlandi} on a different problem. 

\begin{figure}[]
\centering%
\subfigure[]{
\label{fig:p1}
\includegraphics[scale=0.17]{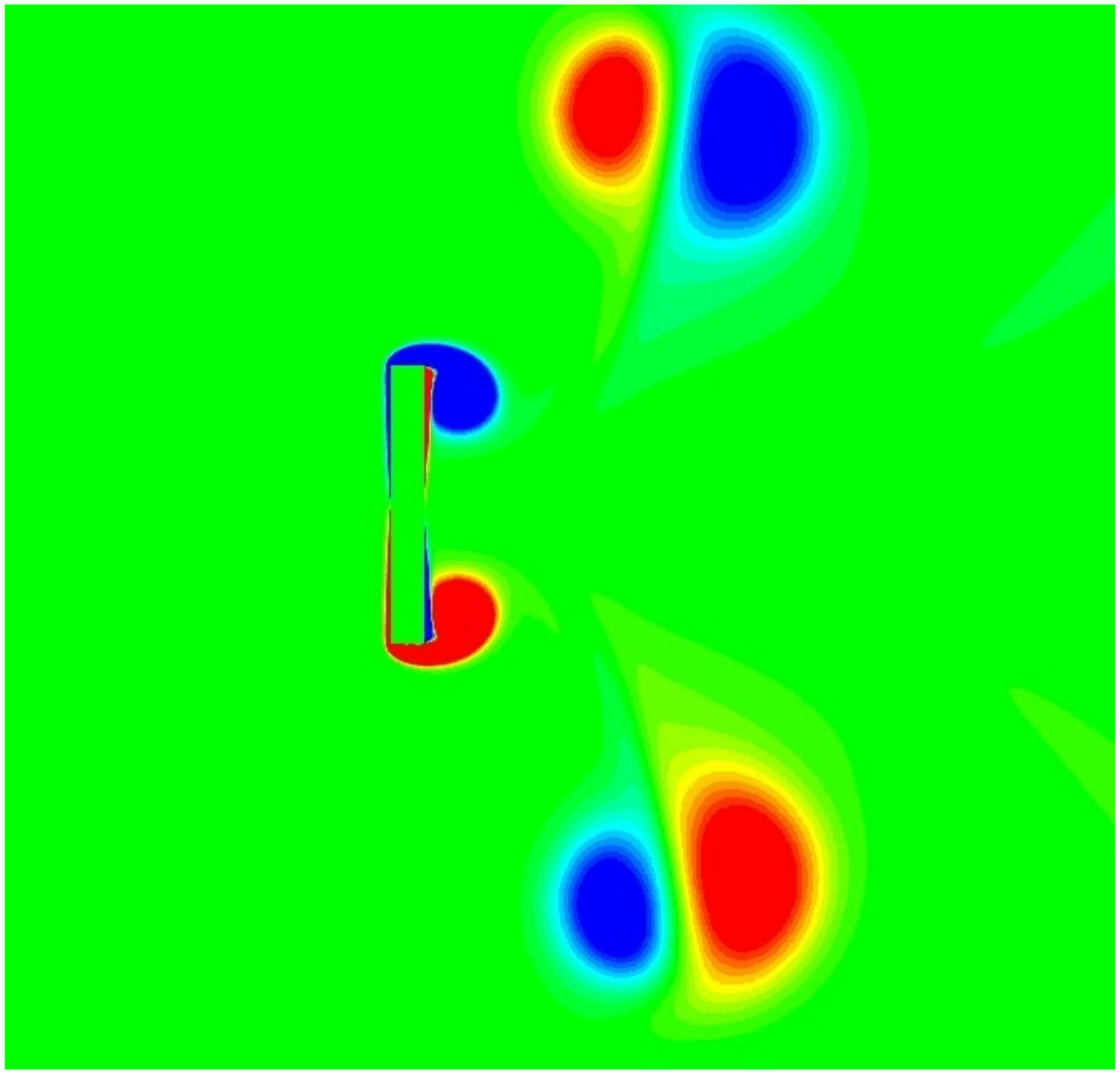}}
\subfigure[]{
\label{fig:p2}
\includegraphics[scale=0.17]{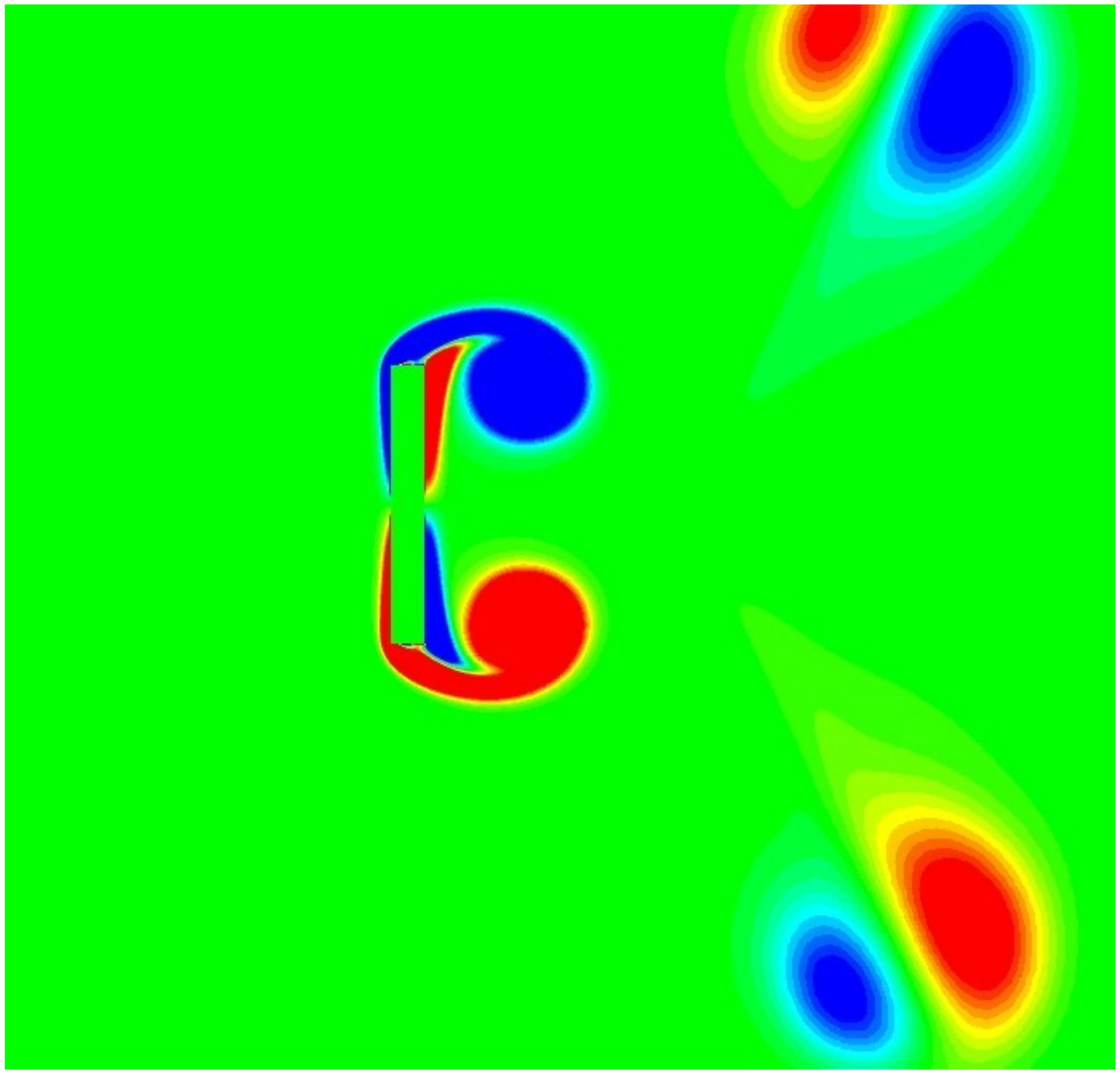}}
\subfigure[]{
\label{fig:p3}
\includegraphics[scale=0.17]{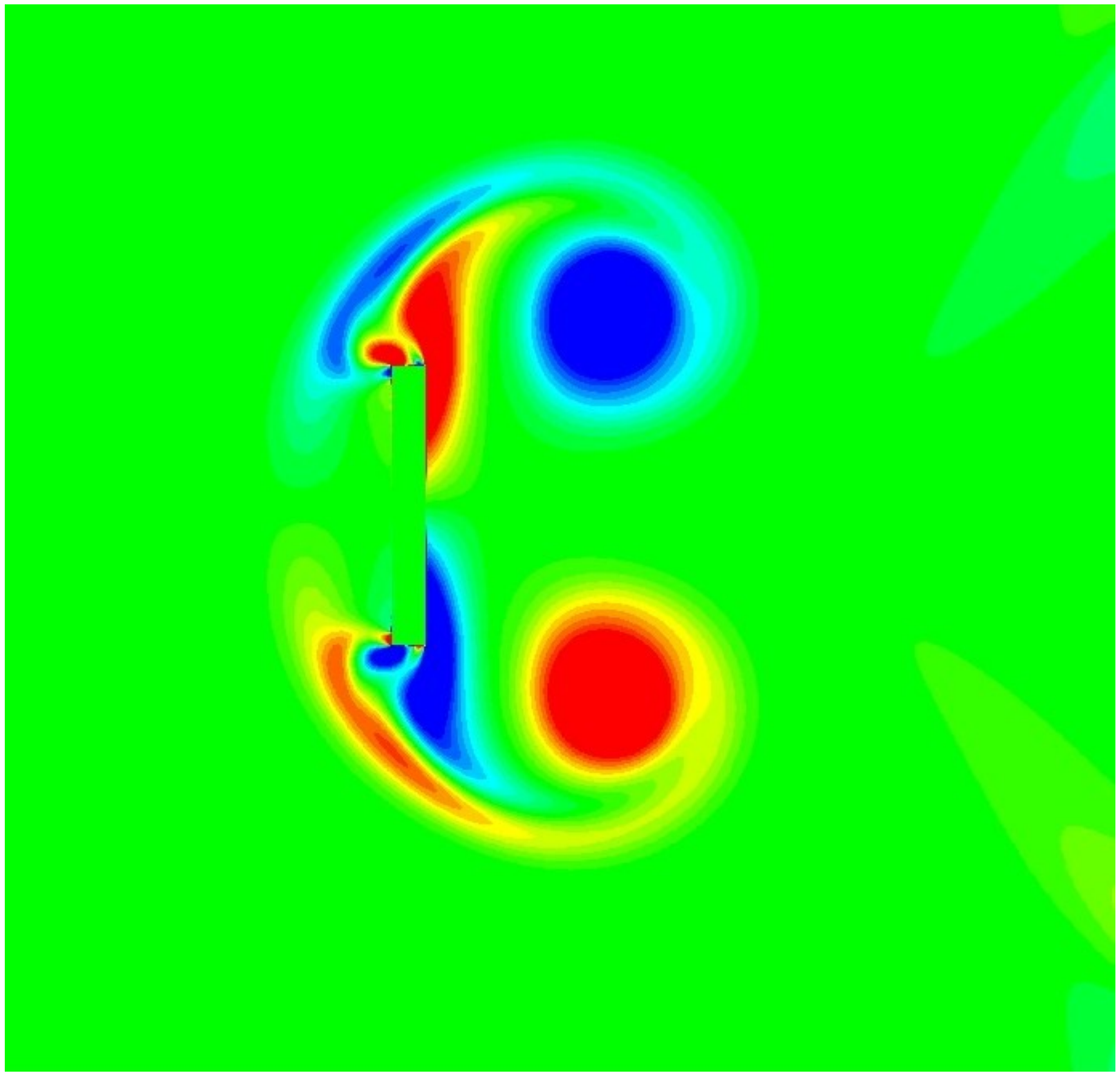}}
\subfigure[]{
\label{fig:p4}
\includegraphics[scale=0.17]{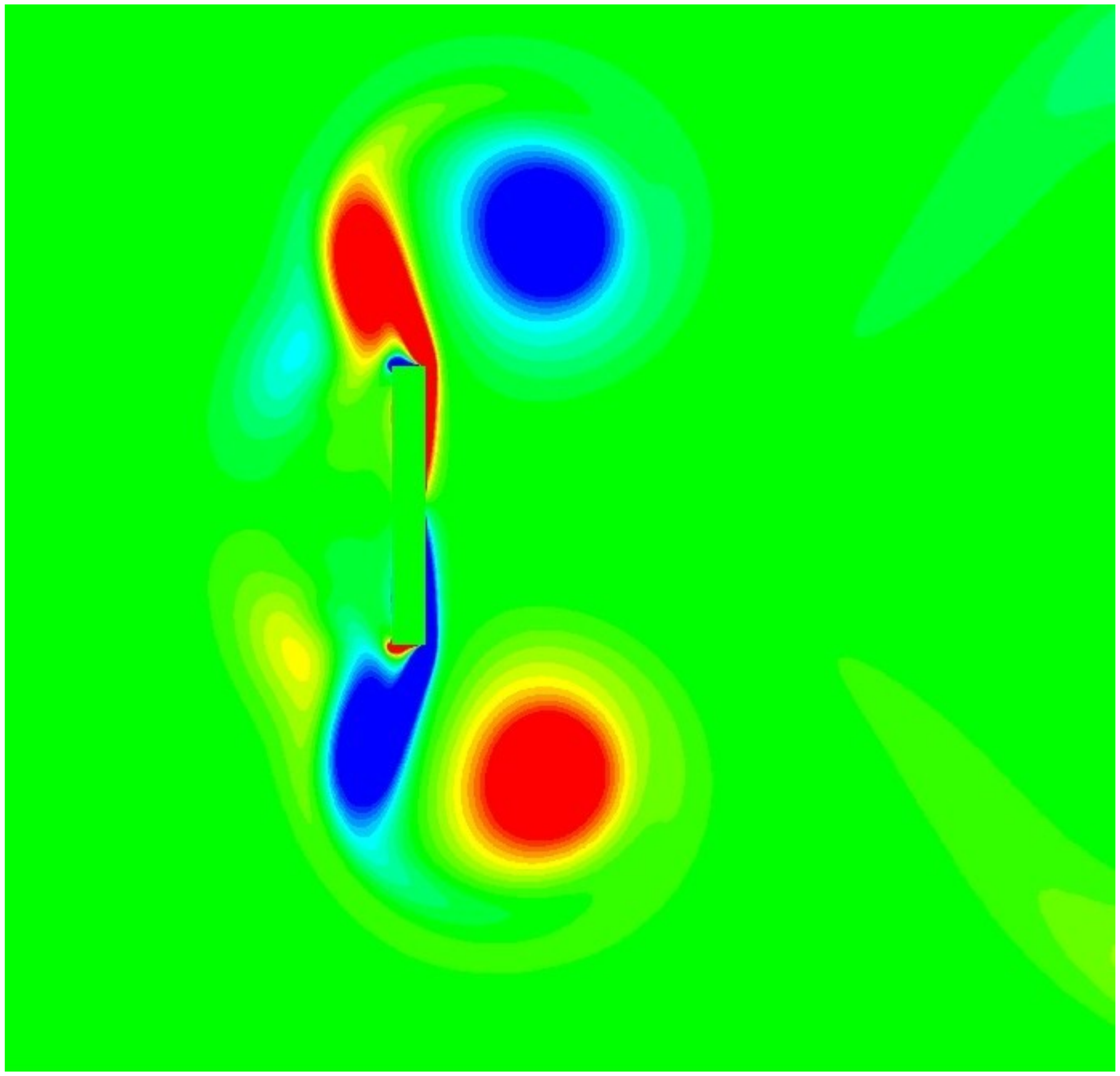}}
\label{fig:allp}
\caption{The S-II mode of vortex shedding; $D/h = 8$, $f_e/f_o = 2$ \& $A/D = 0.175$. Time has been non-dimensionalized using the convective time-scale, $D/U_{\infty}$. (a) Attached primary vortices are growing. (b) Vorticity is generated on the lee side. (c) Close to the cylinder the flow is from right to left. The primary vortices are pushed apart and the secondary vortices are `stretched'. (d) The secondary vortices cut off the supply to primary.}
\end{figure}

This ground effect is clearly visible on a rectangle of aspect ratio 8. The circles in black in figure \ref{fig:s2_phase} indicate the time instances at which vorticity field is plotted in figures \ref{fig:p1} - \ref{fig:p4}. In figure \ref{fig:p1} the cylinder is moving upstream. The primary vortices are seen to form just behind the top and bottom surfaces of the cylinder. Vorticity is continuously supplied to them in the usual manner by the boundary layers. As these primary vortices grow they accelerate the fluid in the wake region leading to the formation of boundary layers, figure \ref{fig:p2}, on the lee side of the cylinder, of oppositely-signed vorticity with respect to the primary vortices. Now, when the cylinder moves downstream there is a local reverse flow near the cylinder. This causes shape changes in both vortices. The secondary vortices continue to grow and cut off supply to the primary vortices as can be seen in Figs. \ref{fig:p3} and \ref{fig:p4}. This cycle repeats.
In the experiments of Xu \textit{et al.} \cite{xu} there was considerable overall reverse flow which aided the formation of opposite signed vortices on the surface of a circular cylinder. Here, aided by the ground effect, we obtain the S-II mode even without reverse flow at the inlet. 

\subsection{Mode competition and chaos}

Perdikaris \textit{et al.} \cite{perdikaris} reported chaotic flow in the wake of a circular cylinder placed in a uniform flow at one particular amplitude of inline oscillation. In their simulations the cylinder was forced at the corresponding Strouhal frequency of the fixed cylinder. At about the same time, without being aware of that work, we had obtained chaotic flow using square and rectangular cylinders. We thus confirm their appealing finding. Further, while they surmised that a competition between antisymmetric and symmetric shedding was causing chaos, they did not have direct evidence for this. In particular, their lift coefficients and spectra indicate antisymmetric shedding under all non-chaotic conditions, so they do not have mode competition between antisymmetric and symmetric modes. The use of a rectangular cross-section makes it easy to obtain symmetric shedding, so we are able to demonstrate that the shedding is antisymmetric at $f_e$ less than for the chaotic flow, and symmetric for $f_e$ greater than this value, which is a direct demonstration of mode competition in the sense of Ciliberto and Gollub \cite{gollub1,gollub2}. 

\begin{figure}[h]
\centering
\subfigure[]{
\label{fig:f2_delay}
\includegraphics[scale=0.17]{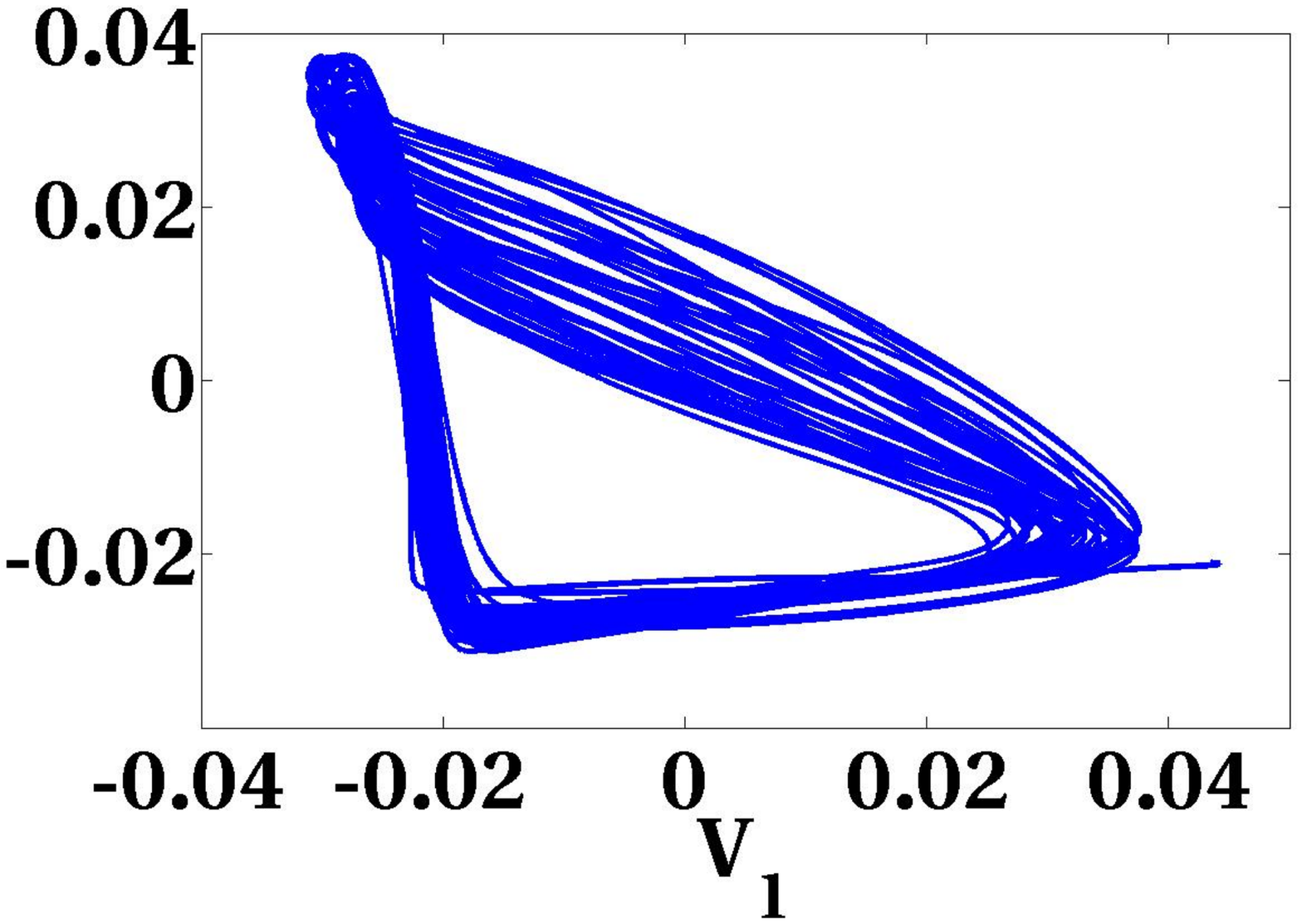}}
\subfigure[]{
\label{fig:f2085_delay}
\includegraphics[scale=0.17]{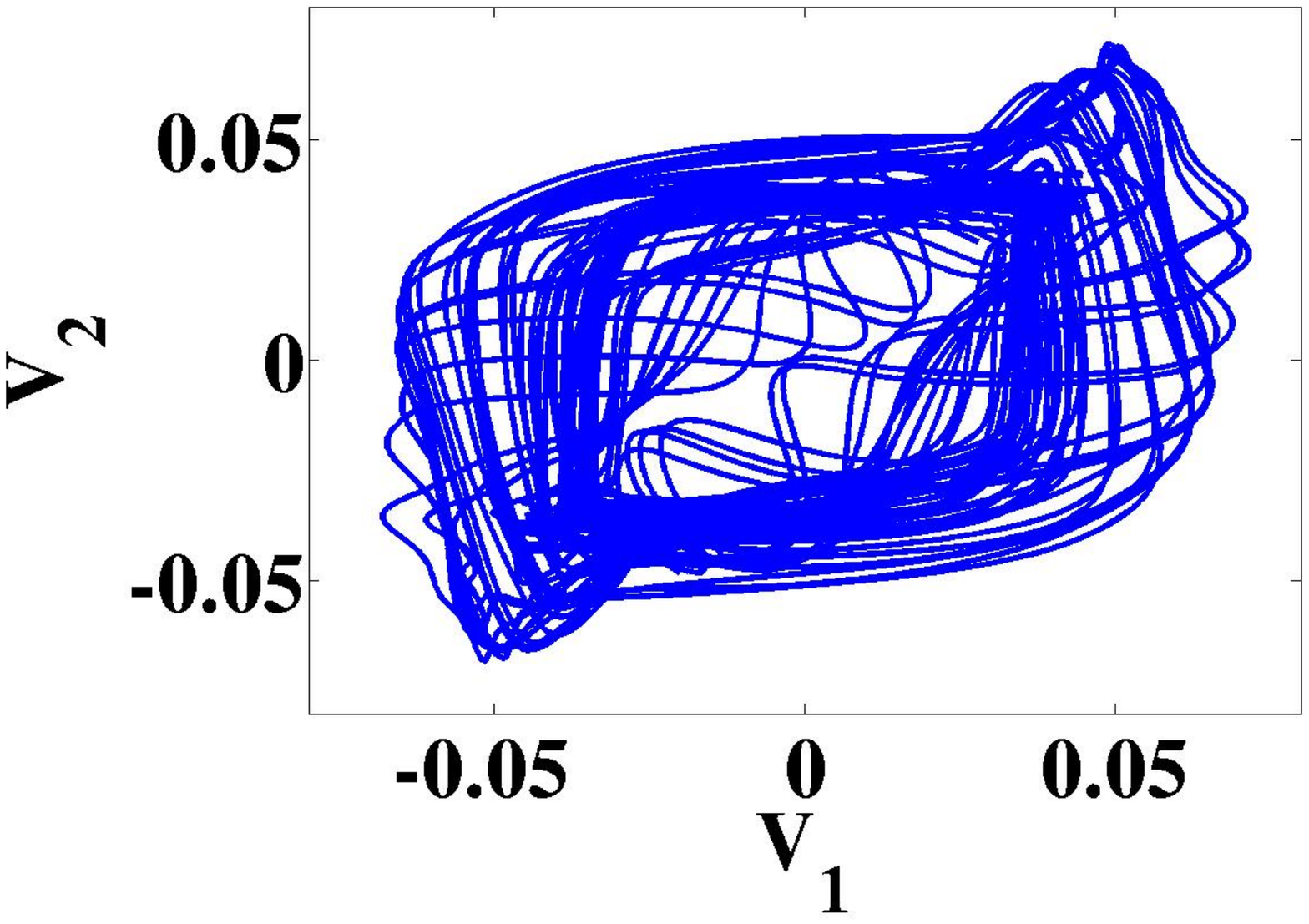}}
\caption{Delay plots for $D/h=4$, $f_e/fo = 2$ and $2.085$. The oscillation amplitude $A/D = 0.175$. The monitor point is behind the cylinder at ($0.5 D, 0.35 D$). The delay plot (a) consists of closed curves, indicating periodicity, whereas (b) is characteristic of an aperiodic time signal.}
\label{fig:f215}
\end{figure}

\begin{figure}
\centering
\subfigure[]{
\label{fig:f2}
\includegraphics[scale=0.18]{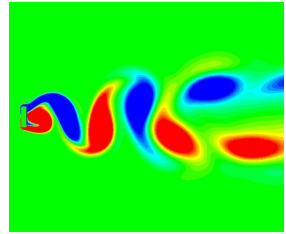}}
\subfigure[]{
\label{fig:f2085}
\includegraphics[scale=0.18]{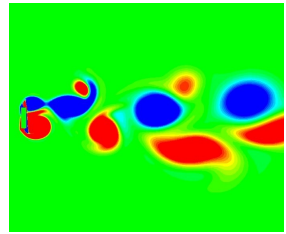}}
\subfigure[]{
\label{fig:f215}
\includegraphics[scale=0.18]{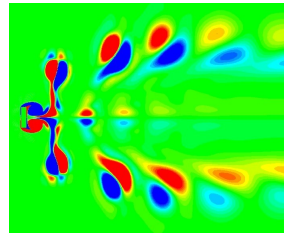}}
\caption{Chaotic window for $D/h = 4$. (a) $f_e/f_o = 2$, the shedding here is antisymmetric. (b) $f_e/f_o = 2.085$, the shedding is chaotic. (c) $f_e/f_o = 2.15$, S-III mode of symmetric shedding.}
\label{fig:chaotic_window}
\end{figure}

In figure \ref{fig:f215}, typical delay plots, for $u_y$ at $f_e/f_o = 2$ and $2.085$ are shown. The axes on the delay plots represent $V_1 = u_y(t + \tau)$ and $V_2 = u_y(t - \tau)$ at a suitably chosen location and delay time $\tau$. In figure \ref{fig:f2_delay} the paths in phase space are closed, which indicates periodicity. The noise in the computations gives rise to a patch rather than a single path, as often happens in these computations. In spite of this noise, this figure may easily be contrasted with figure \ref{fig:f2085_delay}, which is indicative of a chaotic flow. The chaotic window is easily visualised in figure \ref{fig:chaotic_window}, with the antisymmetric mode, locked on to $0.5 f_e$, and the S-III mode, locked on to $f_e$, on either side of the narrow window of chaotic flow in between.

%\begin{figure}[]
%\centering
%\includegraphics[scale=0.135]{f2085.jpg}
%\caption{Vorticity field at a typical time for $D/h=4$, $f_e/f_o$ = 2.085, $A/D = 0.175$. There is no order in the spatial arrangement of shed vortices.}
%\label{fig:energy_compare}
%\end{figure}

\begin{figure}[]
\centering
\includegraphics[scale=0.3]{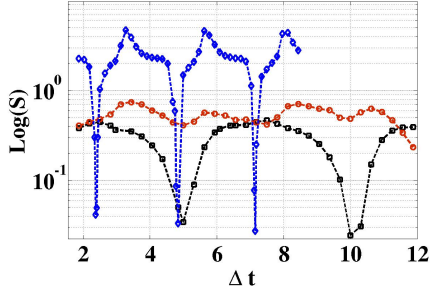}
\caption{$D/h = 4$. Variation of $S$ with $\Delta t$ for different cases. The black curve is for $f_e/f_o = 2$, the red curve for $f_e/f_o = 2.085$ and the blue curve for $f_e/f_o = 2.15$. The black and blue curves exhibit periodicity and have sharp dips at $\Delta t = nT$ for any integer $n$, whereas the red curve is aperiodic.}
\label{fig:aspect4_entropy}
\end{figure}

Thus far we have obtained qualitative indications of the chaotic window. To confirm that the flow is indeed chaotic, we use a global measure $S(\Delta t)$, defined as:
\begin{equation}
S = \sum_i {\sum_j{[\omega(x_i,y_j,t_o + \Delta t) - \omega(x_i,y_j,t_o)]^2} \Delta x \Delta y}.
\end{equation}
Being an integrated quantity over the entire domain, $S$ is a reliable measure of chaos. For a periodic system of period $T$, $S(T)=0$. Moreover for any $\Delta t$ we should have $S(\Delta t+T)=S(\Delta t)$. For $f_e/f_o = 2$ and $f_e/f_o = 2.15$, both these properties are seen in figure \ref{fig:aspect4_entropy}. In particular, a sharp dip in $S$ for $\Delta t = nT$ for any integer $n$ is visible. On the other hand, for $f_e/f_o = 2.085$, $S$ remains at a high value, characterising a chaotic system.

%\begin{figure}[h]
%\centering
%\includegraphics[scale=0.18]{aspect8_entropy.jpg}
%\caption{$D/h = 8$. Variation of $S$ with $\Delta t$ for differnt cases. Black curve: $f_e/f_o = 2$. Maroon curve: $f_e/f_o = 2.5$. Blue curve: $f_e/f_o = 4$. The vorticity field is periodic for $f_e/f_o = 2$. }
%\label{fig:aspect8_entropy}
%\end{figure}
%

\begin{figure}[]
\centering
\subfigure[]{
\label{fig:aspect8_f25_spectrum}
\includegraphics[scale=0.3]{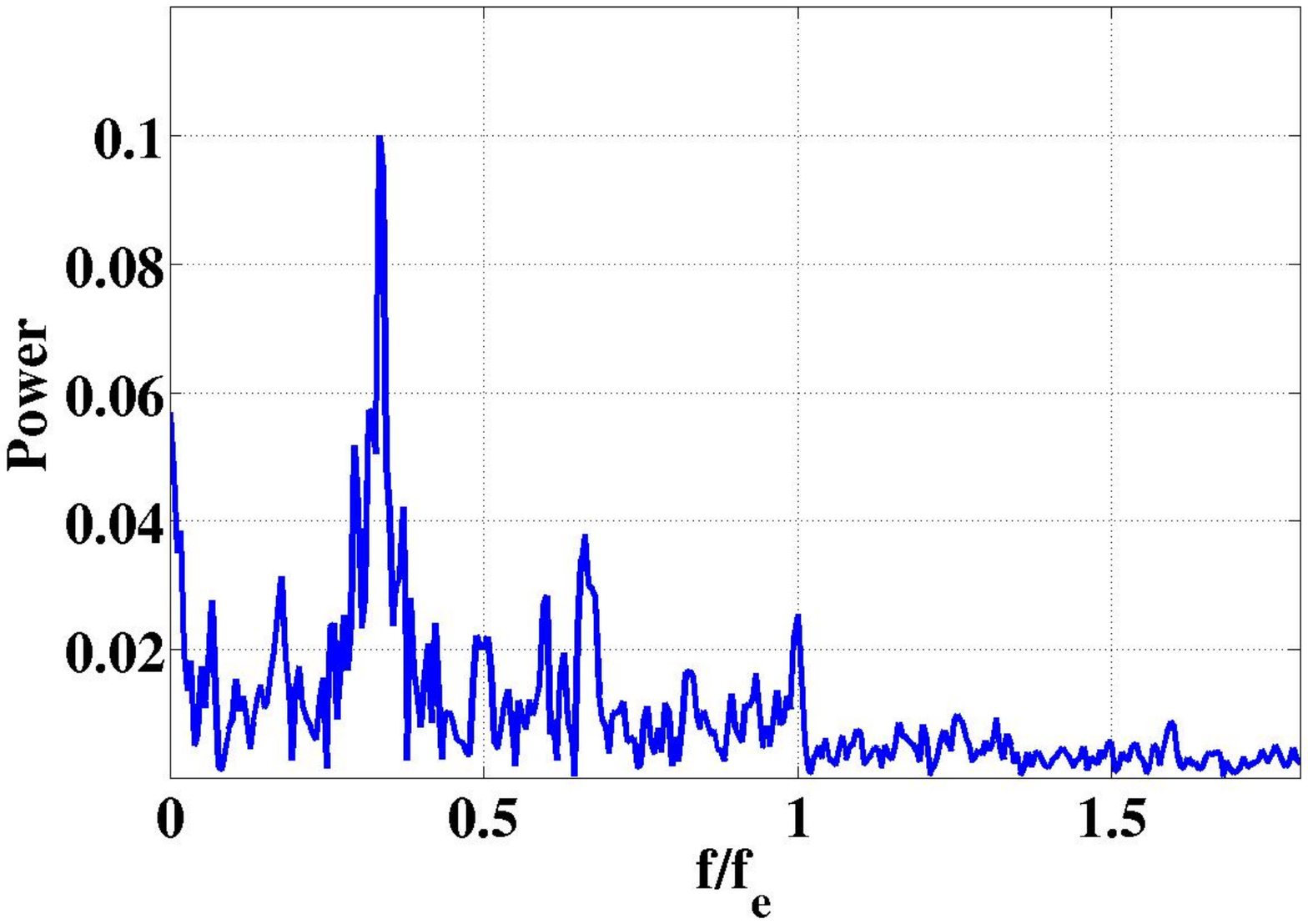}}
\subfigure[]{
\label{fig:aspect8_f25_field}
\includegraphics[scale=0.18]{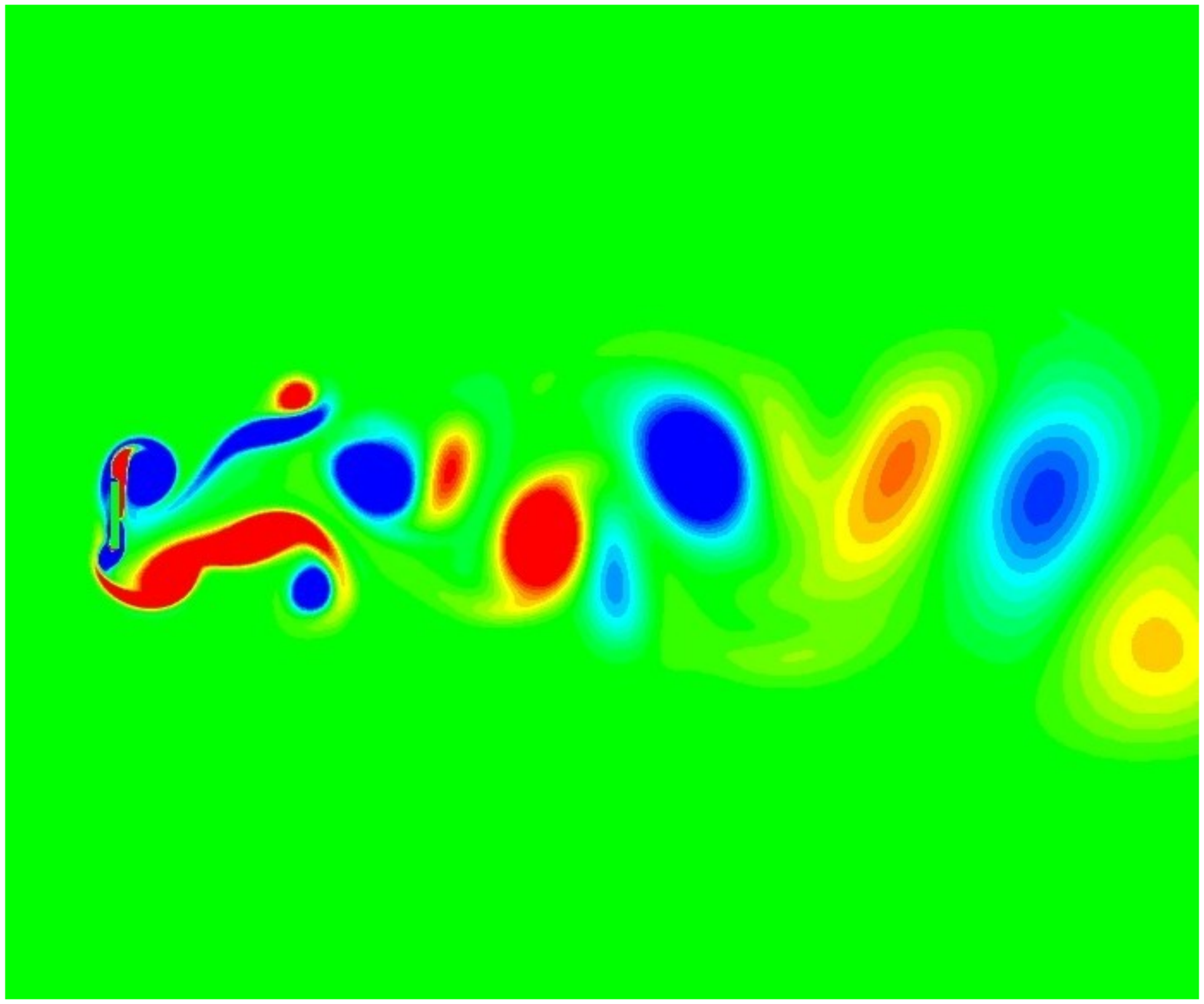}}
\caption{$D/h = 8$, $f_e/f_o$ = 2.5 \& $A/D = 0.175$. (a) The spectrum is broadband, and (b) the  arrangement of vortices is unordered, showing that the flow is chaotic.}
\label{fig:aspect8_chaos}
\end{figure}

\begin{figure}[h]
\centering
\includegraphics[scale=0.3]{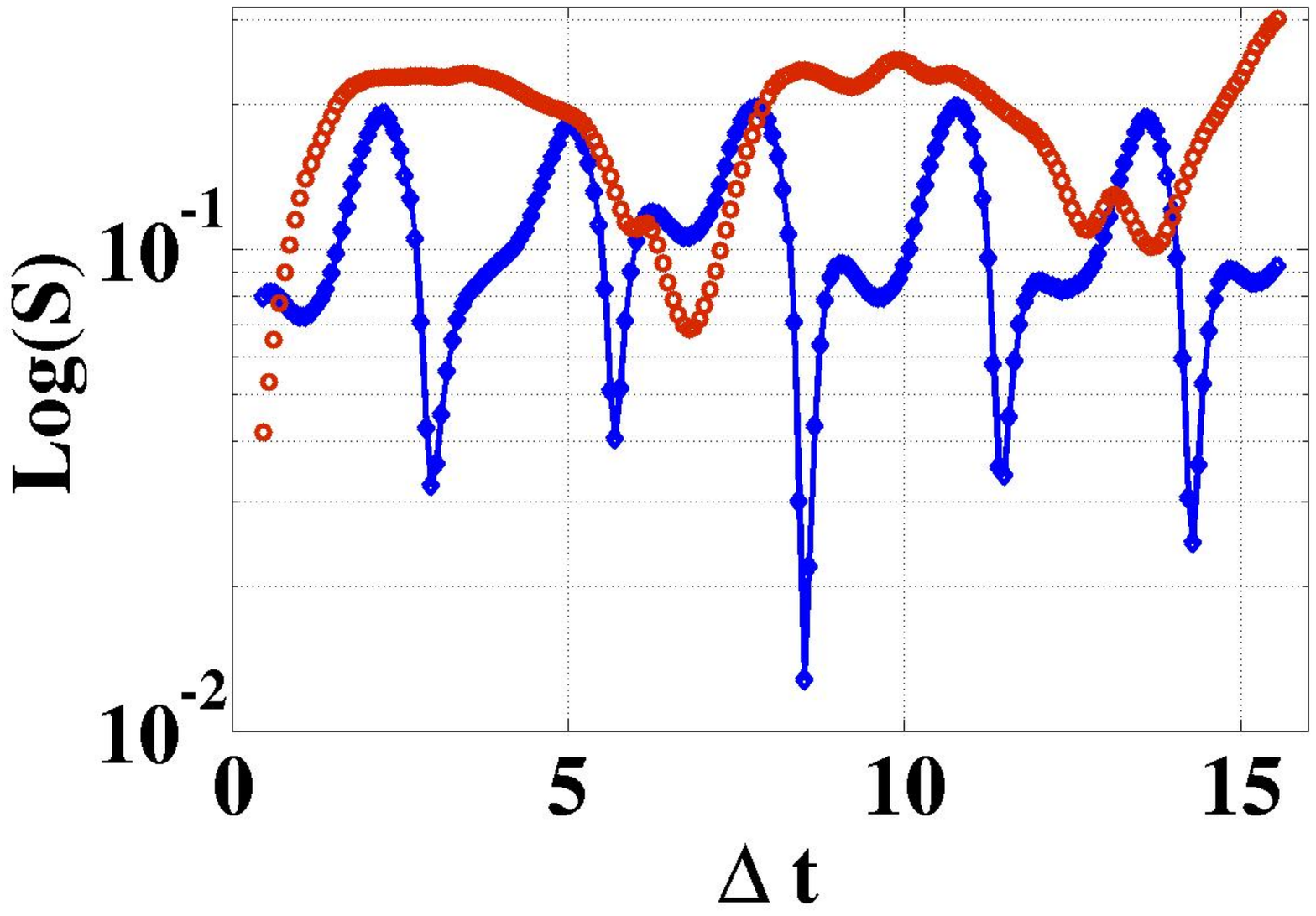}
\caption{$D/h = 8$. Variation of $S$ with $\Delta t$ in a region $x/D = 2.5 - 8.75$. The blue curve is for $f_e/f_o = 2$, and is repetitive indicating periodicity. The red curve is for the case $f_e/f_o = 2.5$, and shows the aperiodic nature of the vorticity field in that region.}
\label{fig:aspect8_entropy_small_region}
\end{figure}

A different behaviour is seen at $D/h = 8$. The vortex shedding mode changes from S-II to the mixed mode with S-I shedding when $f_e/f_o$ is varied from $2$ to $3.5$. Figure \ref{fig:aspect8_entropy_small_region} shows that the flow for $f_e/f_o = 2$ is periodic. However, the flow is chaotic for $f_e/f_o = 2.5$, as evidenced by the spectrum in figure \ref{fig:aspect8_f25_spectrum}. Also the arrangement of vortices at a given time is in no particular pattern, as seen in figure \ref{fig:aspect8_chaos}. The contrast in terms of the $S$ is demonstrated in figure \ref{fig:aspect8_entropy_small_region}. A smaller region is chosen to improve the contrast. The variation of $S$ for $f_e/f_o = 2$ is seen to be periodic, whereas for $f_e/f_o = 2.5$ it is not. 


\subsection{Conclusions}
To summarise, we have studied two-dimensional flow past inline oscillating cylinders of rectangular cross-section at a relatively low Reynolds number. The S-II mode of shedding is obtained computationally for the first time to our knowledge. The S-II mode is enhanced in the case of rectangular cylinders due to the ground effect. A variant symmetric mode, named here as S-III is also observed. The type of shedding changes as the frequency of cylinder oscillation is increased, all other parameters held constant. At lower oscillation amplitudes the frequency regime in between the antisymmetric and symmetric modes consists of a periodic flow where the shedding is neither symmetric nor antisymmetric, but a constant phase is maintained between shedding on the upper and lower surfaces. At higher oscillation frequencies, the flow in this intermediate frequency regime is chaotic. More than one window of chaos may exist in the frequency range. Since these windows always lie between regimes of antisymmetric and symmetric shedding, the chaos is due to mode competition in the sense of Ciliberto \& Gollub \cite{gollub1}. The shedding frequency in the periodic regimes on either side of the chaotic window are locked-on to two different submultiples of the excitation frequency. 

The use of a rectangular cylinder rather than a square one has made several of the above observations possible. A global, and therefore reliable, measure has been used to ascertain the existence of chaos. It is hoped that the present results will motivate experiments with rectangular oscillating cylinders. The work also presents natural extensions to flow past accelerating bodies.

\subsection{Acknowledgements}
The authors would like to thank Prof. Ram Ramaswamy and Dr. Santosh Ansumali for helpful discussions on chaos and the Lattice-Boltzmann method respectively.

\end{document}